\documentclass[a4paper,11pt]{article}
\usepackage{pos}
\usepackage{bm}
\usepackage{braket}

\title{$NJ/\psi$ and $N\eta_c$ interactions from lattice QCD}

\author*[a]{Yan Lyu}
\author[a]{Takumi Doi}
\author[a]{Tetsuo Hatsuda}
\author[b,a]{Takuya Sugiura}
\affiliation[a]{Interdisciplinary Theoretical and Mathematical Sciences Program (iTHEMS), RIKEN, Wako, 351-0198, Japan}

\affiliation[b]{Faculty of Date Science, Rissho University, Kumagaya, 360-0194, Japan}

\emailAdd{yan.lyu@riken.jp}
\emailAdd{doi@ribf.riken.jp}
\emailAdd{thatsuda@riken.jp}
\emailAdd{sugiura@rcnp.osaka-u.ac.jp}

\abstract{The interaction between nucleon and charmonia ($J/\psi$ and $\eta_c$) is expected to deepen our understanding of various aspects in nonperturbative QCD ranging from the origin of nucleon mass to $J/\psi$ mass modification in nuclear medium and properties of hidden-charm pentaquark states. Here, we present the low-energy $NJ/\psi$ and $N\eta_c$ interactions based on ($2+1$) flavor lattice QCD simulations with nearly physical pion mass $m_\pi=146$ MeV. The interactions, extracted from the spacetime correlations of the nucleon and charmonium system by using the HAL QCD method, are found to be attractive in all distances and manifest a characteristic long-range tail consistent with the two-pion exchange interaction.
The resulting scattering lengths are around $0.3$ fm, $0.4$ fm and $0.2$ fm for $NJ/\psi$ with spin $3/2$, with spin $1/2$, and $N\eta_c$, respectively.
Our results are orders of magnitude larger than those from the photoproduction experiments assuming the vector meson dominance.}

\FullConference{%
 The 41st International Symposium on Lattice Field Theory (LATTICE2024)\\
 28 July - 3 August 2024\\
Liverpool, UK\\
}

\begin{document}
\maketitle

\section{Introduction}

The low-energy interaction between a nucleon ($N$) and a charmonium ($J/\psi$ and $\eta_c$)
has deep connections with fundamental questions in QCD, such as how hadrons gain/lose mass in vacuum/medium, what kind of multiquark hadrons can exit.

Firstly, the trace anomaly contribution to the nucleon mass is related to the forward scattering amplitude of $N$ and $J/\psi$. This is because the amplitude involves the nucleon matrix element of gluon field $\langle N|GG|N\rangle$, as shown in Ref.~\cite{Kharzeev:1995ij}. In addition, the
$J/\psi$ mass modification in nuclear medium can be obtained from the $N$-$J/\psi$ scattering length, as discussed in Ref.~\cite{Hayashigaki:1998ey}. Secondly, following the discovery of the first pentaquark state $P_c$ by the LHCb Collaboration in 2015 \cite{LHCb2015_Pc}, a comprehensive understanding of its nature and properties has become a pressing issue. Achieving this goal requires precise knowledge of the $N$-$c\bar c$ interactions, which serve as critical inputs in coupled-channel analyses. Moreover, accurate $N$-$c\bar{c}$ interactions are also crucial for predicting the binding energy of charmonium-nucleus bound states, given large discrepancies remain among various phenomenological methods~\cite{Krein:2017usp}. Thirdly, since $N$ and $c\bar{c}$ do not have common quarks, the interaction between them arises from multiple gluon exchange, which is likely dominated by two-pion exchange at long range~\cite{Fujii:1999xn,Brambilla:2015rqa,Castella2018}. It is important to investigate this hypothesis from lattice QCD.

Experimentally, the low-energy $N$-$c\bar{c}$ interaction is inferred from the $J/\psi$ photonproduction off the proton, which gives rise to a scattering length of $O(1\sim10)\times 10^{-3}$~fm~\cite{Pentchev:2020kao} when assuming the vector meson dominance and $O(1)$ fm~\cite{JPAC:2023qgg} when constrained by the low-energy unitarity. 
Previous lattice claudications~\cite{Yokokawa:2006td,Kawanai:2010ru,Sugiura:2019pye,Liu:2008rza,Skerbis:2018lew} are limited to quenched approximation, heavy pion mass, or large uncertainties.

Given the situation, a realistic lattice QCD study on the low-energy $N$-$c\bar{c}$ interaction has been performed recently in Ref.~\cite{Lyu:2024ttm}, and we report the results in this conference.

\section{Methodology}
The starting point in HAL QCD method~\cite{Ishii2007,Ishii2012} is the following $R$-correlator,
\begin{align}\label{Eq_R}
    R(\bm r,t)&=\sum_{\bm x}\braket{0|N(\bm x+\bm r, t)O_{c\bar c}(\bm x, t)\overline{\mathcal{J}}(0)|0}/e^{-(m_N+m_{c\bar c})t}\\ \nonumber
    &=\sum_{n}a_n\psi_{E_n}(\bm r)e^{-(\Delta E_n)t} + O(e^{-(\Delta E^*)t}),
\end{align}
where $\mathcal{J}(0)$ and $a_n=\braket{N,c\bar c;E_n|\overline{\mathcal{J}}(0)|0}$ are a source operator and the corresponding overlapping factor to $n$-th eigenstate, respectively. 
$\Delta E_n=E_n-(m_N+m_{c\bar c})$ and $\Delta E^*\sim m_\pi$ are the eigenenergy and the inelastic threshold with respect to the $N$-$c\bar c$ threshold.
$R(\bm r, t)$ at large $t$ satisfies the following integrodifferential equation,
\begin{align}
    &\left[\frac{1+3\delta^2}{8\mu}\frac{\partial^2}{\partial t^2}-\frac{\partial}{\partial t} - H_0 +O(\delta^2\partial^3_t)\right] R(\bm r,t)\\ \nonumber
    &=\int d\bm r' U(\bm r, \bm r')R(\bm r',t), \quad \mu=\frac{m_Nm_{c\bar c}}{m_N+m_{c\bar c}}, \quad \delta=\frac{m_N-m_{c\bar c}}{m_N+m_{c\bar c}}.
\end{align}
The nonlocal potetential defined above can be expanded as $U(\bm r,\bm r')=V(r)\delta(\bm r-\bm r')+\sum_{n=1}V_n(\bm r)\nabla^n\delta(\bm r-\bm r')$, leading to a local potential at the leading order, which is accurate for describing near-threshold scattering,
\begin{align}\label{Eq_LO_V}
    V(r)=R^{-1}(\bm r,t)\left[\frac{1+3\delta^2}{8\mu}\frac{\partial^2}{\partial t^2}-\frac{\partial}{\partial t} - H_0 + O(\delta^2\partial^3_t)\right] R(\bm r,t),
\end{align}
where the $O(\delta^2\partial^3_t)$ term is found to be consistent with zero within statistical uncertainties, and is neglected in our study.

\section{Lattice setup}

Local sink operators are adopted in our calculation. They are defined as,
\begin{align}
    N_\alpha(x)  &= \epsilon_{ijk}[u^i(x)C\gamma_5 d^j(x)]u^k_\alpha(x), \label{Eq_N_op} \\ 
    J/\psi_\mu(x) &= \delta_{ij} \bar c^i(x)\gamma_\mu c^j(x), \label{Eq_Jpsi_op} \\ 
    \eta_c(x) &= \delta_{ij} \bar c^i(x)\gamma_5 c^j(x), \label{Eq_etac_op}
\end{align}
with $ijk$ being color indices, $\alpha$ ($\mu$) being spinor (vector) index.
At source, we perform wall-type smearing and  Coulomb gauge fixing.

We use ($2+1$)-flavor lattice QCD configurations generated on a $96^4$ lattice with a spacing of $a \simeq 0.0846$ fm ($a^{-1} \simeq 2333$ MeV), resulting in a physical volume of $La \simeq 8.1$ fm. The Iwasaki gauge action at $\beta = 1.82$ and the nonperturbatively $O(a)$-improved Wilson quark action with stout smearing are employed at nearly physical quark masses~\cite{Ishikawa2016}. For the charm quark, the relativistic heavy quark (RHQ) action is used to eliminate cutoff errors up to next-to-next-to-leading order~\cite{Aoki2003}. Two sets of RHQ parameters (set 1 and set 2)~\cite{Namekawa2017}, tuned close to the physical charm quark mass, allow interpolation ($0.385 \times$ set 1 $+$ $0.615 \times$ set 2) to reproduce the spin-averaged $1S$ charmonium dispersion relation. Shown in Table~\ref{tab-mass} is the mass for relevant hadrons in our study.
\begin{table}[htbp]
\begin{center}
\caption{Hadron masses with statistical errors obtained from lattice QCD together with the experimental values (isospin-averaged).
Two values for $J/\psi$ and $\eta_c$ are from set 1 and set 2 parameters of RHQ action, respectively.
}
\begin{tabular}{ccc}
  \hline\hline
    Hadron &~~Lattice [MeV]  &~~Expt. [MeV] \\
  \hline
    $\pi$ &\ \ 146.4(4)  & \ \ 138.0\\
    $K$  & \ \ 524.7(2) & \ \ 495.6\\
    $N$  &\ \ 954.0(2.9)  & \ \ 938.9 \\
    $J/\psi$ &\ \ 3121.1(1)~~3076.3(1) &\ \ 3096.9\\
    $\eta_c$ &\ \ 3022.8(1)~~2976.3(1)&\ \ 2984.1\\
  \hline\hline
\end{tabular} \label{tab-mass}
\end{center}
\end{table}

\section{Numerical results}
\subsection{Potential}
\begin{figure*}[htbp]
    \centering
   \includegraphics[width=6.0cm]{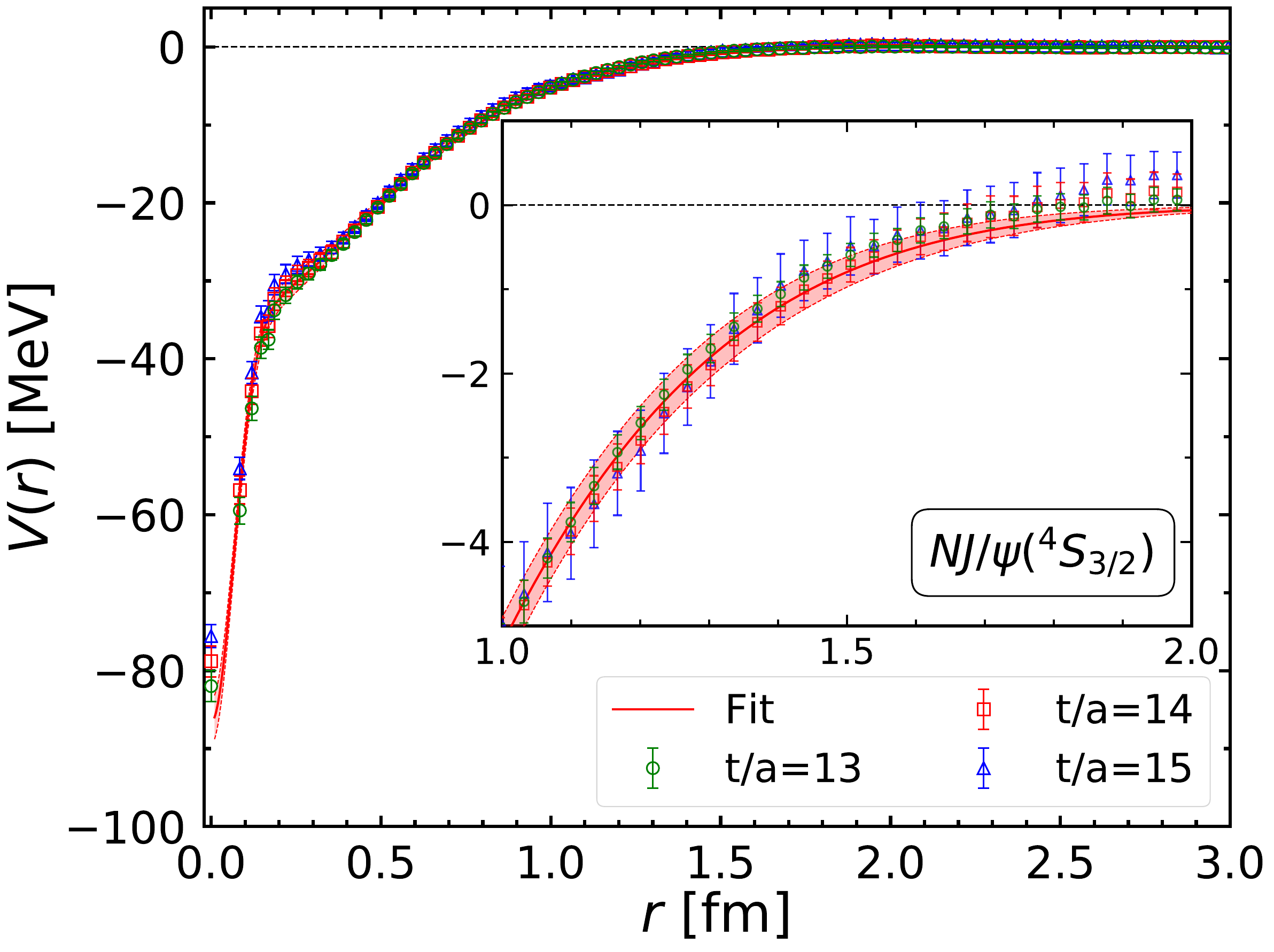}
    \includegraphics[width=6.0cm]{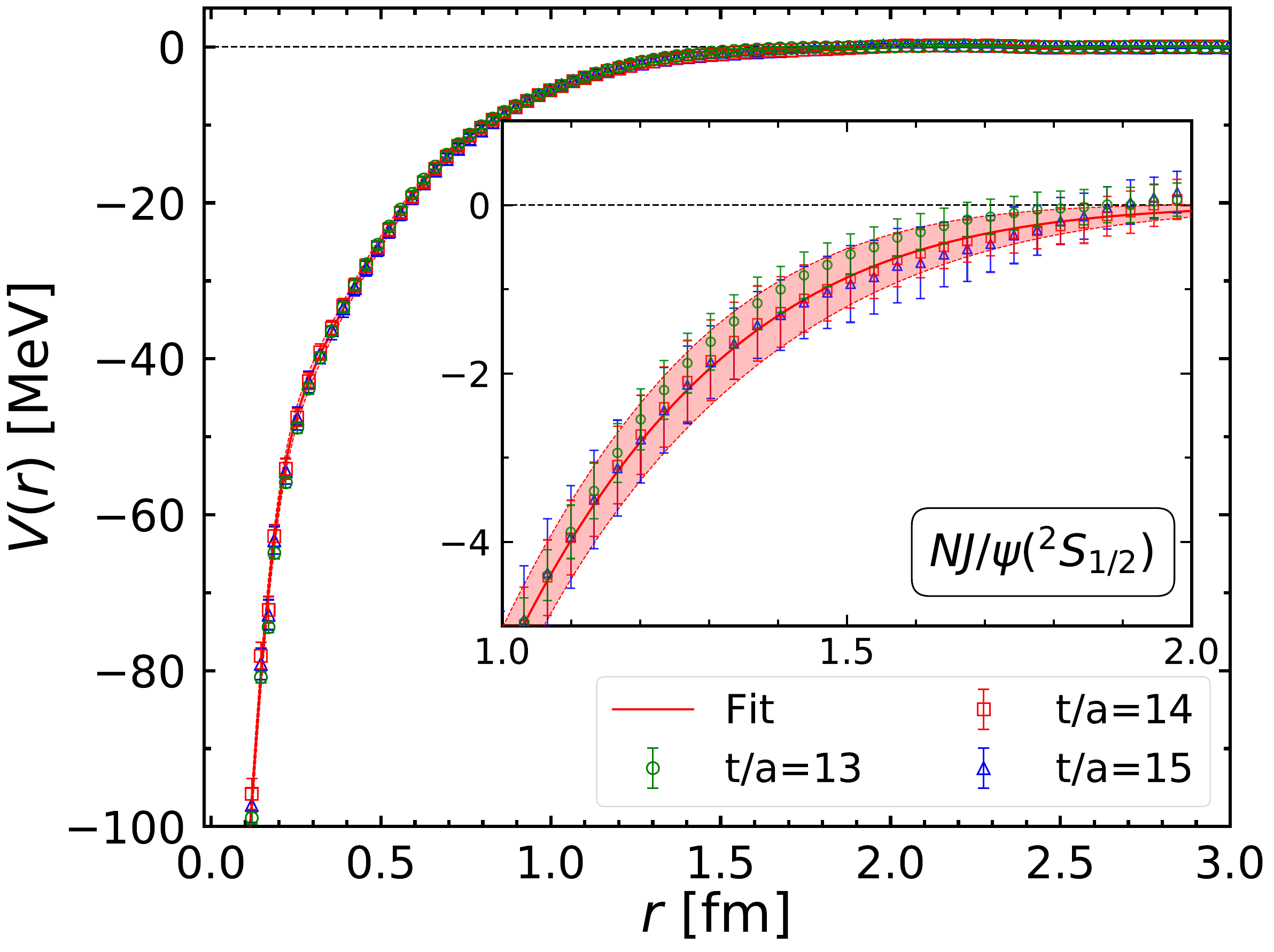}
    \includegraphics[width=6.0cm]{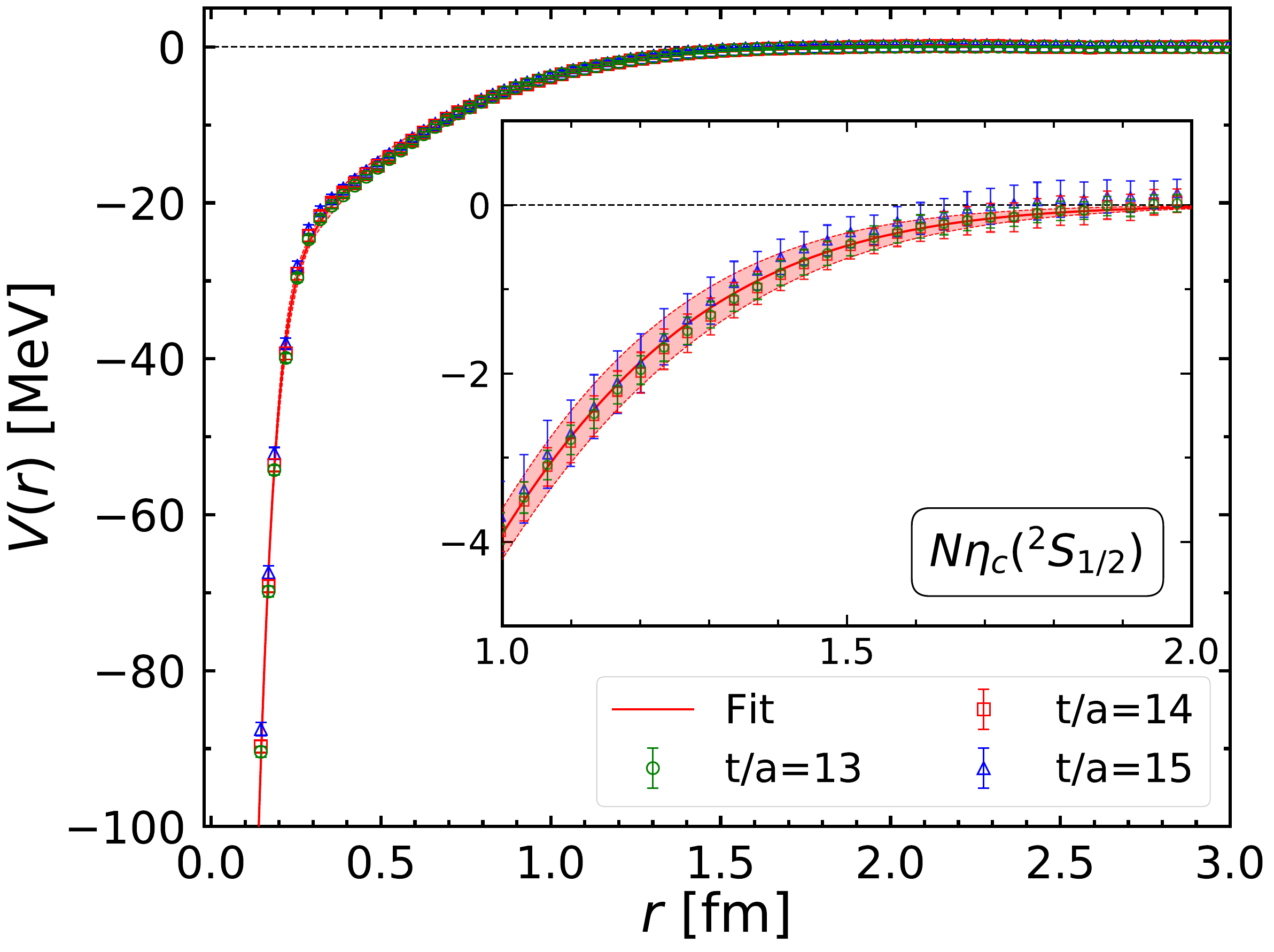}
    \includegraphics[width=6.0cm]{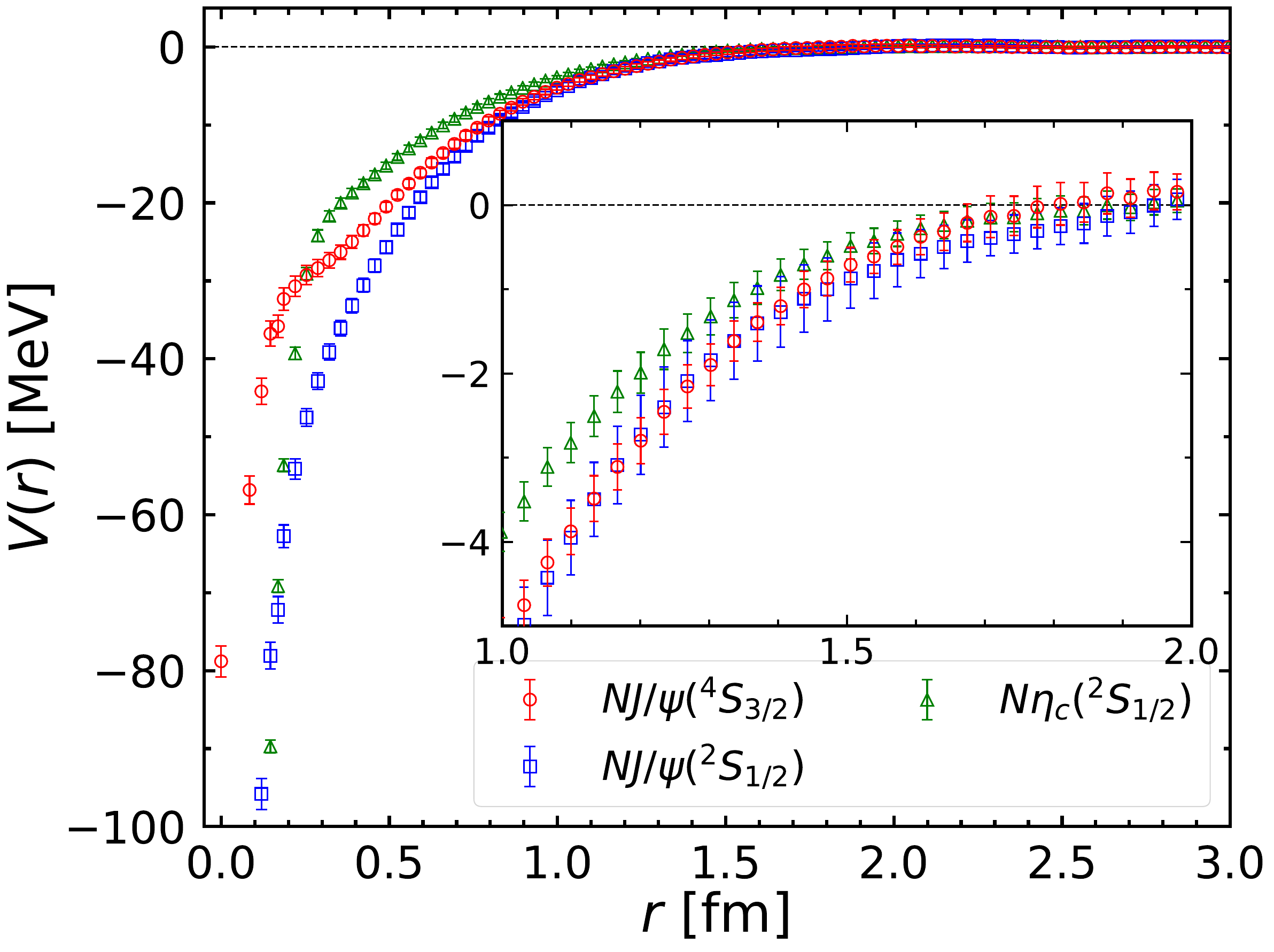}
    \caption{The $N$-$c\bar c$ potential extracted at $t/a=13$, $14$, and $15$ for $N$-$J/\psi$ with $^4S_{3/2}$ (upper left), with $^2S_{1/2}$ (upper right), and $N$-$\eta_c$ with $^2S_{1/2}$ (lower left). The red bands show the fit results with phenomenological three-range Gaussians at $t/a=14$.
    The three potentials at $t/a=14$ are also shown in (lower right) for a direct comparison.
    A magnification is shown in the inset for each panel.
    Figures are taken from \cite{Lyu:2024ttm}.
    }
    \label{Fig-V}
\end{figure*}

In Fig.~\ref{Fig-V}, we show the $N$-$c\bar{c}$ potentials extracted at $t\simeq1.2$~fm.
We find: (i) Potentials show weak $t$ dependence, meaning systematic errors associated with the inelastic states and the truncation of the derivative expansion are small.
(ii) The $N$-$c\bar{c}$ potential is attractive at all distances, qualitatively similar as $N$-$s
\bar{s}$ potential~\cite{Lyu_Nphi_PRD2022}. 
(iii) The long range potentials for all three channels are very similar,
 indicating a common underlying mechanism dictates these three interactions at large distances.

\subsection{Two-pion exchange}
\begin{figure}[htbp]
    \centering
    \includegraphics[width=8.0cm]{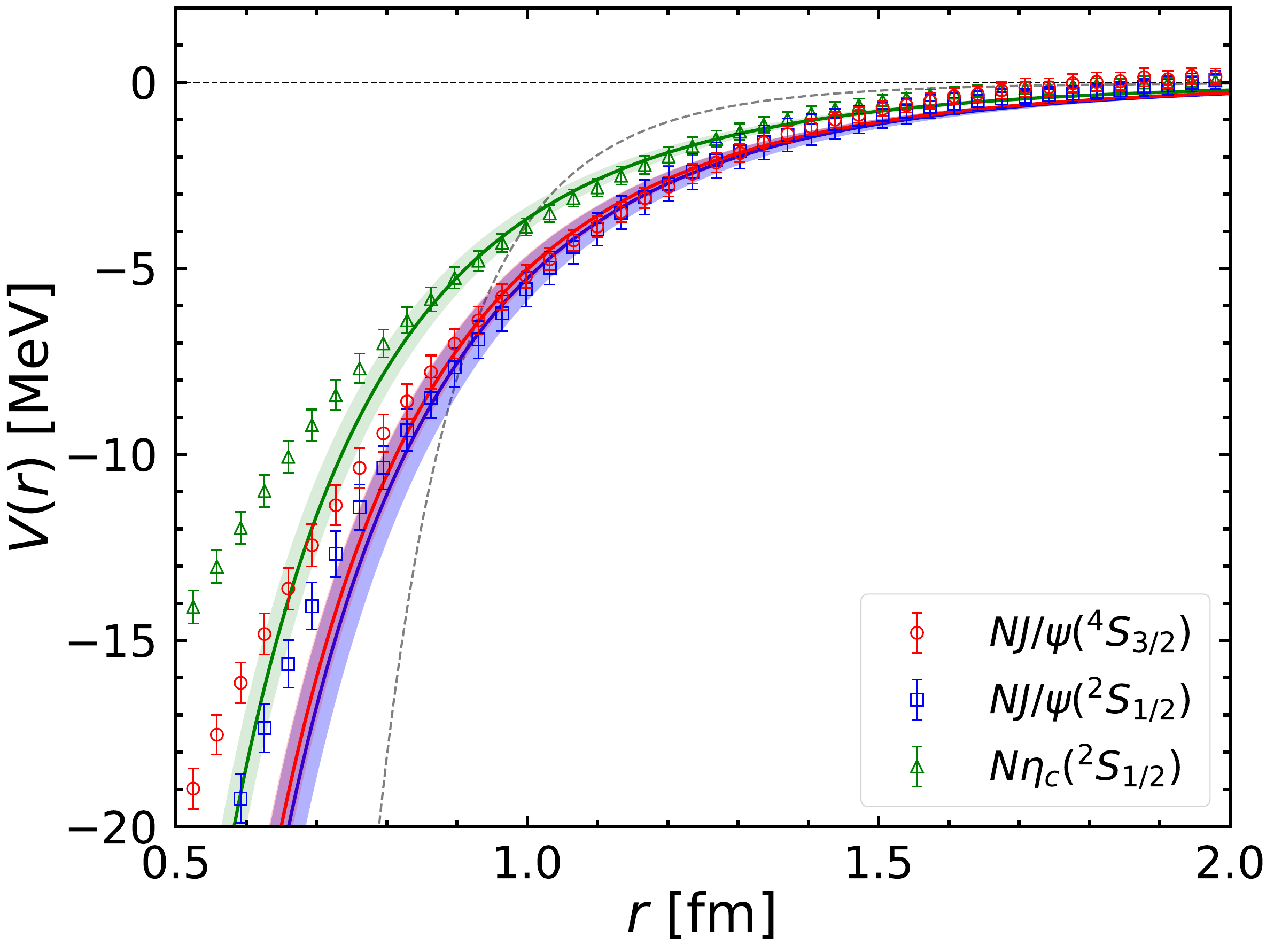}
    \caption{The bands show the fit with the TPE function $V(r)=-\alpha e^{-2m_\pi r}/r^2$ to the long-range $N$-$c\bar{c}$ potentials. The gray dashed line is the best fit with $V(r)=-\alpha/r^7$ for comparison.
    Taken from \cite{Lyu:2024ttm}.
    }
    \label{Fig-V-TPE}
\end{figure}

As discussed before, the potential between $N$ and $c\bar{c}$ initially generated through multiple gluon exchange, is expected to transform into a two-pion exchange mechanism at long distances, which takes the following term in coordinate space $V(r)=-\alpha\frac{e^{-2m_\pi r}}{r^2}$~\cite{Castella2018}.
Theoretically, this potential closely resembles the van der Waals interaction generated from multiple photon exchange in QED, which decreases in power-law in coordinate space $V(r)=-\alpha/r^7$.
Shown in Fig.~\ref{Fig-V-TPE} is the best fit to the long range potential ($0.9< r< 1.8~\text{fm}$) by $V(r)=\alpha\frac{e^{-2m_\pi r}}{r^2}$ (bands) with a fit parameter $\alpha=22(2)$, $23(3)$, and $16(2)$ $\text{MeV}\text{fm}^2$ for $N$-$J/\psi$ with spin $3/2$, with spin $1/2$, and $N$-$\eta_c$, respectively. 
This result implies that the long-range $N$-$c\bar{c}$ potentials are consistent with the two-pion exchange interaction.

\subsection{Phase shifts and scattering parameters}
\begin{figure}[htbp]
    \centering
    \includegraphics[width=8.0cm]{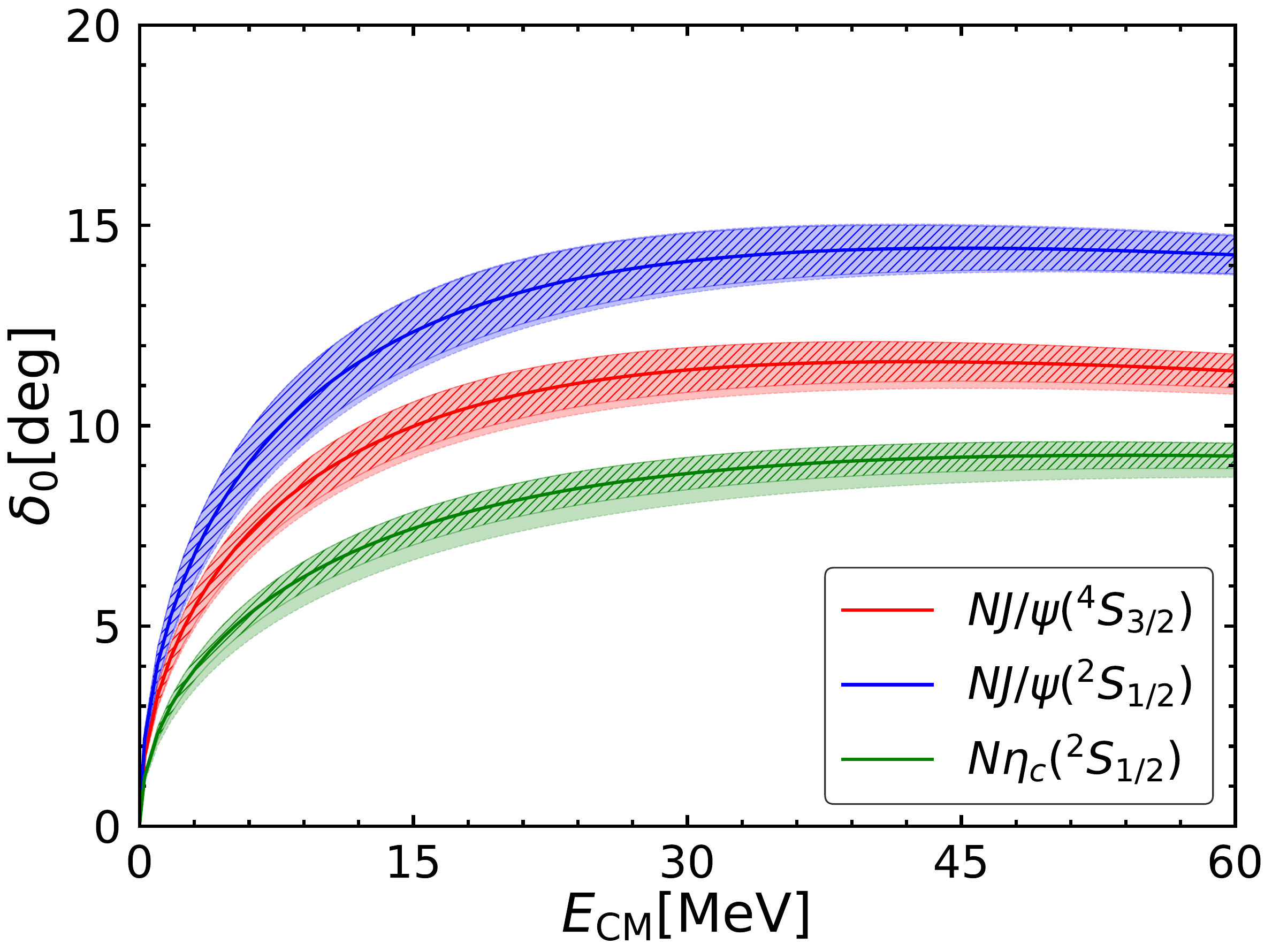}
    \caption{The $N$-$c\bar c$ scattering phase shifts. The central value and statistical error obtained with $V_\text{fit}(r)$ at $t/a=14$ are shown by solid lines and inner bands. The outer bands show total uncertainty obtained by adding statistical error and systematic error in quadrature.
    Taken from \cite{Lyu:2024ttm}.
    }
    \label{Fig-delta}
\end{figure}

In order to convert the lattice potential to physical observables, we fit the potentials in Fig.~\ref{Fig-V} with three-range Gaussians $V_\text{fit}(r)=-\sum_{i=1}^3 a_ie^{(-r/b_i)^2}$, and corresponding fitting results are shown by red bands in Fig.~\ref{Fig-V}.
The scattering phase shits are calculated with $V_\text{fit}(r)$ and shown in Fig.~\ref{Fig-delta}.
Scattering length ($a_0$) and effective range ($r_\text{eff}$) are extracted from the near threshold phase shifts through $k\cot\delta_0=\frac{1}{a_0}+\frac12 r_\text{eff}k^2+O(k^4)$, and tabulated in Table~\ref{tab-scattering}.

\begin{table}[htbp]
\begin{center}
\caption{The $N$-$c\bar c$ scattering length $a_0$ and effective range $r_\text{eff}$ with statistical error ($1$st parentheses) and systematic error ($2$nd parentheses).}
\begin{tabular}{lll}
  \hline\hline
    ~~~channel~~~~~~~~&~~~$a_0$ [fm]~~~~~~~~~&~~~$r_\text{eff}$ [fm]\\
  \hline
$N$-$J/\psi(^4S_{3/2})$ ~~~~~~~~~&$0.30(2)\left(^{+0}_{-2}\right)$ ~~~~~~~~~~&$3.25(12)\left(^{+6}_{-9}\right)$ \\
$N$-$J/\psi(^2S_{1/2})$ ~~~~~~~~~&$0.38(4)\left(^{+0}_{-3}\right)$ ~~~~~~~~~~&$2.66(21)\left(^{+0}_{-10}\right)$\\
$N\eta_c(^2S_{1/2})$ ~~~~~~~~~&$0.21(2)\left(^{+0}_{-1}\right)$ ~~~~~~~~~~&$3.65(19)\left(^{+0}_{-6}\right)$ \\
\hline\hline
\end{tabular} \label{tab-scattering}
\end{center}
\end{table}   

Using the scattering length in Table~\ref{tab-scattering}, we are able to predict the $J/\psi$ mass reduction in normal nuclear medium with density $\rho_\text{nm}=0.17~\text{fm}^{-3}$,
\begin{align}
    \delta m_{J/\psi}\simeq \frac{2\pi(m_N+ m_{J/\psi})}{m_Nm_{J/\psi}}a^\text{spin-av}_{J/\psi} \rho_\text{nm}=19(3)~\text{MeV},
\end{align}
where the $N$-$J/\psi$ spin-averaged scattering length is defined as $a^\text{spin-av}_{NJ/\psi}=\frac{2a^\text{(3/2)}_0+a^\text{1/2)}_0}{3}$.

In Table~\ref{comp}, we summarize the spin-averaged $N$-$J/\psi$ scattering length in the literature, which involves various approaches including: photoproduction reaction, QCD multipole expansion, QCD sum rule, coupled channel analysis, and lattice QCD.
Previous studies in one way or another surfer from either badly fixed parameters in corresponding models
or lacking of some important dynamics of QCD in corresponding calculations, which also highlights the importance of performing realistic lattice QCD calculations as what we did in the present study.

\begin{table}[htbp]
\begin{center}
\caption{The spin-averaged $N$-$J/\psi$ scattering length in the literature compared with the result from the present study shown in the last line.
}
\begin{tabular}{cccc}
\hline \hline
$a^\text{spin-av}_{NJ/\psi}$ [fm] & Year & Author & Method \\
\hline
0.046(5) & 2016 & Gryniuk-Vanderhaeghen~\cite{Gryniuk:2016mpk} & Photoproduction (VMD) \\
$3 \sim 25 \cdot 10^{-3}$ & 2021 & Pentchev-Strakovsky~\cite{Pentchev:2020kao} & Photoproduction (VMD) \\
$O(1)$ & 2023 & JPAC~\cite{JPAC:2023qgg} & Photoproduction (unitarity) \\
0.05 & 1992 & Kaidalov-Volkovitsky~\cite{Kaidalov:1992hd} & QCD multipole expansion \\
0.24 & 1997 & Brodsky-Miller~\cite{Brodsky:1997gh}& QCD multipole expansion  \\
$\geq 0.37$ & 2005 & Sibirtsev-Voloshin~\cite{Sibirtsev:2005ex} & QCD multipole expansion \\
$0.2 \sim 3 \cdot 10^{-3}$ & 2020 & Du-Baru-Guo et. al.~\cite {Du:2020bqj} & Coupled channel \\
0.10(2) & 1999 & Hayashigaki \cite{Hayashigaki:1998ey}& QCD sum rule \\
0.71(48) & 2006 & Yokokawa-Sasaki et al. \cite{Yokokawa:2006td}& LQCD (quenched) \\
0.1(7) & 2008 & Liu-Lin-Orginos \cite{Liu:2008rza}& LQCD (full, extrapolate to phys. pt.) \\
0.33(5) & 2010 & Kawanai-Sasaki \cite{Kawanai:2010ru}& LQCD (quenched) \\
0.47(1) & 2019 & Sugiura-Ikeda-Ishii \cite{Sugiura:2019pye}& LQCD ($m_{\pi} = 700$ Mev) \\
$\approx 0$ & 2019 & Skerbis-Prelovsek \cite{Skerbis:2018lew}& LQCD ($m_{\pi} = 266$ Mev) \\
\textbf{0.33(4)} & 2024 & \textbf{The present work} & LQCD ($m_{\pi} = 146$ Mev) \\
\hline\hline
\end{tabular}\label{comp}
\end{center}
\end{table}   

\section{Summary}

We present a realistic study on the low-energy $N$-$c\bar{c}$ interaction based on ($2+1$)-flavor lattice QCD simulations with nearly physical light quark masses ($m_\pi=146$ MeV) and physical charm quark mass.
The potential derived by the HAL QCD method is attractive for all distances, and possesses a characteristic long-range behavior consistent with the two-pion exchange interaction. 
The resulting phase shifts and scattering parameters are shown in Fig.~\ref{Fig-delta} and Table~\ref{tab-scattering}.

We are under way to perform physical-point calculations with the ($2+1$)-flavor lattice QCD configurations generated by the HAL QCD Collaboration~\cite{Aoyama:2024cko}.
Also, we plan to investigate the universality of the long-range potential across other hadron pairs.

\acknowledgments
We thank members of the HAL QCD Collaboration for stimulating discussions.
We thank members of the PACS Collaboration for the gauge configuration generation conducted on the K computer at RIKEN.
The lattice QCD measurements have been performed on Fugaku and HOKUSAI supercomputers at RIKEN.
We thank ILDG/JLDG \cite{ldg}, which serves as essential infrastructure in this study.
This work was partially supported by HPCI System Research Project (
hp230075, hp230207, hp240157 and hp240213), the JSPS (Grant Nos. JP18H05236, JP22H00129, JP19K03879, JP21K03555, and JP23H05439),   
``Program for Promoting Researches on the Supercomputer Fugaku'' (Simulation for basic science: from fundamental laws of particles to creation of nuclei) 
and (Simulation for basic science: approaching the new quantum era)
(Grants No. JPMXP1020200105, JPMXP1020230411), 
and Joint Institute for Computational Fundamental Science (JICFuS).
YL, TD and TH weresupported by Japan Science and Technology Agency (JST) as part of Adopting Sustainable Partnerships for Innovative Research Ecosystem (ASPIRE), Grant Number JPMJAP2318.

\ifnum  1>0

\fi


\begin{thebibliography}{99}

\bibitem{Kharzeev:1995ij}
  D.~Kharzeev, \emph{Quarkonium interactions in QCD}, 
   \href {https://doi.org/10.3254/978-1-61499-215-8-105} {{Proc. Int. Sch. Phys. Fermi \textbf{130} (1996) 105}} [\href {http://arxiv.org/abs/nucl-th/9601029}{\tt{nucl-th/9601029}}].

 \bibitem{Hayashigaki:1998ey}
  A.~Hayashigaki, \emph{\ensuremath{J/\psi} nucleon scattering length and in-medium mass shift of \ensuremath{J/\psi} in QCD sum rule analysis}, \href {https://doi.org/10.1143/PTP.101.923} {Prog. Theor. Phys. \textbf{101} (1999) 923}
  [\href {http://arxiv.org/abs/nucl-th/9811092} {\tt{nucl-th/9811092}}].

  \bibitem{LHCb2015_Pc}
  R.~Aaij, et~al., \emph{Observation of $j/\ensuremath{\psi}p$ resonances consistent with pentaquark states in ${\mathrm{\ensuremath{\Lambda}}}_{b}^{0}\ensuremath{\rightarrow}j/\ensuremath{\psi}{K}^{\ensuremath{-}}p$ decays}, \href {https://doi.org/10.1103/PhysRevLett.115.072001}{Phys. Rev. Lett. \textbf{115} (2015) 072001} [\href{https://arxiv.org/abs/1507.03414}{\tt{hep-ex/1507.03414}}]


  \bibitem{Krein:2017usp}
  G.~Krein, A.~W. Thomas, K.~Tsushima,  \emph{Nuclear-bound quarkonia and heavy-flavor hadrons}, \href {https://doi.org/10.1016/j.ppnp.2018.02.001}{Prog. Part. Nucl. Phys. \textbf{100} (2018) 161}
  [\href {http://arxiv.org/abs/1706.02688} {\tt{hep-ph/1706.02688}}].

  \bibitem{Fujii:1999xn}
  H.~Fujii, D.~Kharzeev, \emph{Long range forces of QCD}, \href{https://doi.org/10.1103/PhysRevD.60.114039}{Phys. Rev. D \textbf{60} (1999) 114039} [\href{https://doi.org/10.48550/arXiv.hep-ph/9903495}{\tt{hep-ph/9903495}}].
  
  \bibitem{Brambilla:2015rqa}
  N.~Brambilla, G.~a. Krein, J.~Tarr\'us~Castell\`a, A.~Vairo, \emph{Long-range properties of $1S$ bottomonium states}, \href {https://doi.org/10.1103/PhysRevD.93.054002}{Phys. Rev. D \textbf{93} (2016) 054002}
  [\href {http://arxiv.org/abs/1510.05895} {\tt{hep-ph/1510.05895}}].
  
  \bibitem{Castella2018}
  J.~Tarr\'us~Castell\`a, G.~a. Krein, \emph{Effective field theory for the nucleon-quarkonium interaction}, \href {https://doi.org/10.1103/PhysRevD.98.014029} {Phys. Rev. D \textbf{98} (2018) 014029}
  [\href {http://arxiv.org/abs/1803.05412} {\tt{hep-ph/1803.05412}}].

  \bibitem{Pentchev:2020kao}
  L.~Pentchev, I.~I. Strakovsky, \emph{$J/\psi$-$p$ Scattering Length from the Total and Differential Photoproduction Cross Sections},  \href {https://doi.org/10.1140/epja/s10050-021-00364-4}{Eur. Phys. J. A \textbf{57} (2021) 56}
  [\href {http://arxiv.org/abs/2009.04502} {\tt{hep-ph/2009.04502}}].
  
  \bibitem{JPAC:2023qgg}
  D.~Winney, et~al., \emph{Dynamics in near-threshold \ensuremath{J/\psi} photoproduction}, \href {https://doi.org/10.1103/PhysRevD.108.054018}{Phys. Rev. D \textbf{108} (2023) 054018}
  [\href {http://arxiv.org/abs/2305.01449} {\tt{hep-ph/2305.01449}}].

  \bibitem{Yokokawa:2006td}
  K.~Yokokawa, S.~Sasaki, T.~Hatsuda, A.~Hayashigaki, \emph{First lattice study of low-energy charmonium-hadron interaction}, \href {https://doi.org/10.1103/PhysRevD.74.034504} {Phys. Rev. D \textbf{74} (2006) 034504}
  [ \href {http://arxiv.org/abs/hep-lat/0605009} {\tt{hep-lat/0605009}}].
  
 \bibitem{Kawanai:2010ru}
  T.~Kawanai, S.~Sasaki, \emph{Charmonium-nucleon interaction from lattice QCD with a relativistic heavy quark action}, \href{https://doi.org/10.22323/1.105.0156}{PoS LATTICE2010 \textbf{2010} (2010)156} [\href{https://doi.org/10.48550/arXiv.1011.1322}{\tt{hep-lat/1011.1322}}].

\bibitem{Sugiura:2019pye}
  T.~Sugiura, Y.~Ikeda, N.~Ishii, \emph{Charmonium-nucleon interactions from $2+1$ flavor lattice QCD}, \href{https://doi.org/10.22323/1.334.0093}{PoS LATTICE2018 \textbf{2018} (2019) 093} [\href{https://doi.org/10.48550/arXiv.1905.02336}{\tt{hep-lat/1905.02336}}].

\bibitem{Liu:2008rza}
  L.~Liu, H.-W. Lin, K.~Orginos, \emph{Charmed Hadron Interactions}, \href{https://doi.org/10.22323/1.066.0112}{PoS LATTICE2008 \textbf{2008} (2008) 112} [\href{https://doi.org/10.48550/arXiv.0810.5412}{\tt{hep-lat/0810.5412}}].

\bibitem{Skerbis:2018lew}
  U.~Skerbis, S.~Prelovsek, \emph{Nucleon-$J/\psi$ and nucleon-$\eta_{c}$ scattering in $P_{c}$ pentaquark channels from LQCD}, \href{https://doi.org/10.1103/PhysRevD.99.094505}{Phys. Rev. D \textbf{99} (2019) 094505} [\href{https://doi.org/10.48550/arXiv.1811.02285}{\tt{hep-lat/1811.02285}}].

\bibitem{Lyu:2024ttm}
  Y.~Lyu, T.~Doi, T.~Hatsuda, T.~Sugiura, \emph{Nucleon-charmonium interactions from lattice QCD}, \href{https://doi.org/10.1016/j.physletb.2024.139178}{Phys. Lett. B \textbf{860} (2025) 139178} [\href{https://doi.org/10.48550/arXiv.2410.22755}{\tt{hep-lat/2410.22755}}].

\bibitem{Ishii2007}
  N.~Ishii, S.~Aoki, T.~Hatsuda, \emph{Nuclear Force from Lattice QCD}, \href{https://doi.org/10.1103/PhysRevLett.99.022001}{Phys. Rev. Lett. \textbf{99} (2007) 022001}  [\href{https://doi.org/10.48550/arXiv.0611096}{\tt{nucl-th/0611096}} ]
  
\bibitem{Ishii2012}
  N.~Ishii, S.~Aoki, T.~Doi, T.~Hatsuda, Y.~Ikeda, T.~Inoue, K.~Murano, H.~Nemura, K.~Sasaki, \emph{Hadron-hadron interactions from imaginary-time Nambu-Bethe-Salpeter wave function on the lattice}, \href{https://doi.org/10.1016/j.physletb.2012.04.076}{Phys. Lett. B \textbf{712} (2012) 437} [\href{https://doi.org/10.48550/arXiv.1203.3642}{\tt{hep-lat/1203.3642}}].

\bibitem{Ishikawa2016}
  K.-I. Ishikawa, N.~Ishizuka, Y.~Kuramashi, Y.~Nakamura, Y.~Namekawa, Y.~Taniguchi, N.~Ukita, T.~Yamazaki, T.~Yoshie, \emph{2+1 Flavor QCD Simulation on a $96^4$ Lattice}, \href{https://doi.org/10.22323/1.251.0075}{PoS LATTICE2015 \textbf{2015} (2016) 075 }[\href{https://doi.org/10.48550/arXiv.1511.09222}{\tt{hep-lat/1511.09222}}].

\bibitem{Aoki2003}
  S.~Aoki, Y.~Kuramashi, S.-i. Tominaga, \emph{Relativistic heavy quarks on the lattice}, \href{https://doi.org/10.1143/PTP.109.383}{Prog. Theor. Phys. \textbf{109} (2003) 383 }[\href{https://doi.org/10.48550/arXiv.hep-lat/0107009}{\tt{hep-lat/0107009}}].

\bibitem{Namekawa2017}
  Y.~Namekawa, \emph{Charm physics by $N_f=2+1$ Iwasaki gauge and the six stout smeared $O(a)$-improved Wilson quark actions on a $96^4$ lattice}, \href{https://doi.org/10.22323/1.256.0125}{PoS LATTICE2016 \textbf{2016} (2017) 125 }[\href{https://doi.org/10.48550/arXiv.1701.00102}{\tt{hep-lat/1701.00102}}].

\bibitem{Ishikawa2023}
  K.-I. Ishikawa, I.~Kanamori, H.~Matsufuru, I.~Miyoshi, Y.~Mukai, Y.~Nakamura, K.~Nitadori, M.~Tsuji, \emph{102 PFLOPS lattice QCD quark solver on Fugaku}, \href{https://doi.org/10.1016/j.cpc.2022.108510}{Comput. Phys. Commun. \textbf{282} (2023) 108510 }[\href{https://doi.org/10.48550/arXiv.2109.10687}{\tt{hep-lat/2109.10687}}].

\bibitem{Lyu_Nphi_PRD2022}
  Y.~Lyu, T.~Doi, T.~Hatsuda, Y.~Ikeda, J.~Meng, K.~Sasaki, T.~Sugiura, \emph{Attractive $N$-$\phi$ interaction and two-pion tail from lattice QCD near physical point}, \href{https://doi.org/10.1103/PhysRevD.106.074507}{Phys. Rev. D \textbf{106} (2022) 074507 }[\href{https://doi.org/10.48550/arXiv.2207.11186}{\tt{hep-lat/2207.11186}}].

\bibitem{Gryniuk:2016mpk}
  O.~Gryniuk, M.~Vanderhaeghen, \emph{Accessing the real part of the forward $J/\psi$-$p$ scattering amplitude from $J/\psi$ photoproduction on protons around threshold}, \href{https://doi.org/10.1103/PhysRevD.94.074001}{Phys. Rev. D \textbf{94} (2016) 074001} [\href{https://doi.org/10.48550/arXiv.1608.08205}{\tt{hep-ph/1608.08205}}].

\bibitem{Kaidalov:1992hd}
  A.~B. Kaidalov, P.~E. Volkovitsky, \emph{Heavy quarkonia interactions with nucleons and nuclei}, \href{https://doi.org/10.1103/PhysRevLett.69.3155}{Phys. Rev. Lett. \textbf{69} (1992) 3155} [\href{https://doi.org/10.48550/arXiv.hep-ph/9209201}{\tt{hep-ph/9209201}}].

\bibitem{Brodsky:1997gh}
  S.~J. Brodsky, G.~A. Miller, \emph{Is $J/\psi$ - nucleon scattering dominated by the gluonic van der Waals interaction?}, \href{https://doi.org/10.1016/S0370-2693(97)01045-9}{Phys. Lett. B \textbf{412} (1997) 125} [\href{https://doi.org/10.48550/arXiv.hep-ph/9707382}{\tt{hep-ph/9707382}}].

\bibitem{Sibirtsev:2005ex}
  A.~Sibirtsev, M.~B. Voloshin, \emph{The Interaction of slow $J/\psi$ and $\psi'$ with nucleons}, \href{https://doi.org/10.1103/PhysRevD.71.076005}{Phys. Rev. D \textbf{71} (2005) 076005} [\href{https://doi.org/10.48550/arXiv.hep-ph/0502068}{\tt{hep-ph/0502068}}].

\bibitem{Du:2020bqj}
  M.-L. Du, V.~Baru, F.-K. Guo, C.~Hanhart, U.-G. Mei\ss{}ner, A.~Nefediev, I.~Strakovsky, \emph{Deciphering the mechanism of near-threshold $J/\psi$ photoproduction}, \href{https://doi.org/10.1140/epjc/s10052-020-08620-5}{Eur. Phys. J. C \textbf{80} (2020) 1053} [\href{https://doi.org/10.48550/arXiv.2009.08345}{\tt{hep-ph/2009.08345}}].

\bibitem{Aoyama:2024cko}
  T.~Aoyama, T.~M. Doi, T.~Doi, E.~Itou, Y.~Lyu, K.~Murakami, T.~Sugiura, \emph{Scale setting and hadronic properties in light quark sector with $(2+1)$-flavor Wilson fermions at the physical point}, \href{https://doi.org/10.1103/PhysRevD.110.094502}{Phys. Rev. D \textbf{110} (2024)094502} [\href{https://doi.org/10.48550/arXiv.2406.16665}{\tt{hep-lat/2406.16665}}].

\bibitem{ldg}
  \url{http://www.lqcd.org/ildg} and \url{http://www.jldg.org}.
 

\end{thebibliography}
\end{document}